\documentclass[fleqn,10pt]{wlscirep}
\usepackage[utf8]{inputenc}
\usepackage[T1]{fontenc}

\usepackage{soul} 
\usepackage{caption}
\usepackage{subcaption}
\usepackage{booktabs}
\usepackage{graphicx}
\usepackage{array}
\usepackage{url}
\usepackage{xcolor}
\usepackage{hyperref}
\newcolumntype{P}[1]{>{\centering\arraybackslash}p{#1}}

\title{Exploration of Interpretability Techniques for Deep COVID-19 Classification using Chest X-ray Images}

\author[1,2,3*+]{Soumick Chatterjee}
\author[4,5+]{Fatima Saad}
\author[4,5,6+]{Chompunuch Sarasaen}
\author[2,7+]{Suhita Ghosh}
\author[2,7]{Valerie Krug}
\author[8,9]{Rupali Khatun}
\author[10]{Rahul Mishra}
\author[11]{Nirja Desai}
\author[8,12]{Petia Radeva}
\author[4,5,13]{Georg Rose}
\author[2,7]{Sebastian Stober}
\author[5,6,13,14]{Oliver Speck}
\author[1,2,13]{Andreas N{\"u}rnberger}

\affil[1]{Data and Knowledge Engineering Group, Otto von Guericke University Magdeburg, Germany}
\affil[2]{Faculty of Computer Science, Otto von Guericke University Magdeburg, Germany}
\affil[3]{Genomics Research Centre, Human Technopole, Italy}
\affil[4]{Institute for Medical Engineering, Otto von Guericke University Magdeburg, Germany}
\affil[5]{Research Campus STIMULATE, Otto von Guericke University Magdeburg, Germany}
\affil[6]{Biomedical Magnetic Resonance, Otto von Guericke University Magdeburg, Germany}
\affil[7]{Artificial Intelligence Lab, Otto von Guericke University Magdeburg, Germany}
\affil[8]{Department of Mathematics and Computer Science, University of Barcelona, Barcelona, Spain}
\affil[9]{Translational Radiobiology, Department of Radiation Oncology, Universitätsklinikum Erlangen, Erlangen, Germany}
\affil[10]{Apollo Hospitals, Bilaspur, India}
\affil[11]{HCG Cancer Centre, Vadodara, India}
\affil[12]{Computer Vision Centre, Cerdanyola, Barcelona, Spain}
\affil[13]{Centre for Behavioural Brain Sciences, Magdeburg, Germany}
\affil[14]{German Centre for Neurodegenerative Diseases, Magdeburg, Germany}

\affil[*]{soumick.chatterjee@ovgu.de / contact@soumick.com}
\affil[+]{these authors contributed equally to this work}

\keywords{COVID-19, Pneumonia, Chest X-ray, Multilabel Image Classification, Deep Learning, Model Ensemble, Interpretability Analysis}

\begin{abstract}
The outbreak of COVID-19 has shocked the entire world with its fairly rapid spread and has challenged different sectors. One of the most effective ways to limit its spread is the early and accurate diagnosing infected patients. Medical imaging, such as X-ray and Computed Tomography (CT), combined with the potential of Artificial Intelligence (AI), plays an essential role in supporting medical personnel in the diagnosis process. Thus, in this article five different deep learning models (ResNet18, ResNet34, InceptionV3, InceptionResNetV2 and DenseNet161) and their ensemble, using majority voting have been used to classify COVID-19, pneumoni{\ae} and healthy subjects using chest X-ray images. Multilabel classification was performed to predict multiple pathologies for each patient, if present. Firstly, the interpretability of each of the networks was thoroughly studied using local interpretability methods - occlusion, saliency, input X gradient, guided backpropagation, integrated gradients, and DeepLIFT, and using a global technique - neuron activation profiles. The mean Micro-F1 score of the models for COVID-19 classifications ranges from 0.66 to 0.875, and is 0.89 for the ensemble of the network models. The qualitative results showed that the ResNets were the most interpretable models. This research demonstrates the importance of using interpretability methods to compare different models before making a decision regarding the best performing model.
\end{abstract}
\begin{document}

\flushbottom
\maketitle
%
%


\section*{Introduction}
\label{sec1}
In 2020, the world witnessed a serious new global health crisis: the outbreak of the infectious COVID-19 disease, which is caused by the Severe Acute Respiratory Syndrome Coronavirus 2 (SARS-CoV-2)~\cite{zhu2020novel,li2020early}. Due to its long incubation period and its highly contagious nature, it is important to identify infected cases early and isolate them from the healthy population. To date, viral nucleic acid detection using Reverse Transcription Polymerase Chain Reaction (RT-PCR) has been regarded as the gold standard diagnostic method~\cite{c19}. However, RT-PCR tests have been reported to suffer from a high rate of false negatives owing to laboratory and sample collection errors~\cite{ai2020correlation,fang2020sensitivity}. 

However, medical imaging emerges as a great alternative candidate for screening COVID-19 cases and discriminating them from other conditions, as the majority of infected patients exhibit abnormalities on medical chest imaging~\cite{zhang2020covid,narin2020automatic,apostolopoulos2020covid}. In this context, chest radiography (CXR) and Computed Tomography (CT) are widely utilised in front-line hospitals for diagnosis~\cite{kanne2020chest,bernheim2020chest,xie2020chest}. In certain instances, chest CT images have been demonstrated to exhibit higher sensitivity than RT-PCR and have detected COVID-19 infections in patients with negative RT-PCR results~\cite{xie2020chest,huang2020use,omer2020covid,ai2020correlation}. 
Nevertheless, there are numerous advantages to encourage the use of CXR imaging in clinical practice, such as faster diagnosis, infection control, and lesser harmfulness than CT~\cite{rubin2020role,harahwa2020optimal}. Moreover, X-ray machines are far more readily available than CT scanners, especially in developing countries. In addition, with the help of portable X-ray machines, imaging can be performed in the isolation rooms, decreasing the risk of infection transmission during transportation to the CT room, as well as the time needed for disinfecting the CT equipment and room~\cite{jacobi2020portable}. Despite its limitations, CXR is more widely available than CT across the globe and is widely utilised for COVID-19 screening~\cite{jacobi2020portable}.

Airspace opacities or ground-glass opacities (GGO) are commonly reported radiological appearances with COVID-19~\cite{guan2020clinical, durrani2020chest}. The predominant distributions in the bilateral, peripheral, and lower zones are primarily observed (90\%)~\cite{wong2020frequency}. However, these manifestations are very similar to various viral pneumoni{\ae} and other inflammatory and infectious lung diseases. Therefore, it is difficult for radiologists to discriminate COVID-19 from other types of pneumoni{\ae}~\cite{ng2020imaging}. Expert radiologists are needed to achieve high diagnostic performance, and the duration of the diagnostic is relatively long. 

Artificial intelligence (AI) can play one of the potential roles in strengthening the power of imaging tools to provide accurate diagnosis. Many AI applications have focused on infection quantification and identification to assist radiologists in decision-making. The classification of COVID-19 and other types of pneumonia has been investigated using deep learning techniques~\cite{zhang2020covid, cohen2020covid}. However, due to the "black box" nature, the rationale behind such techniques is often unknown; hence, these techniques are considered to have low reliability to be integrated within the clinical workflow. Interpretability techniques, which show the focus area of such deep learning methods, are potentially needed to build the confidence of medical practitioners in such methods. Techniques have been proposed that also involve interpretability to understand the reasoning performed by the model~\cite{ozturk2020automated}. However, comparative studies of different models based on accuracy and interpretability, and then verification of the interabilities by doctors have not been performed. Thereby, in this work, the authors have considered the state-of-the-art deep learning models to classify COVID-19 and similar pathologies, along with a thorough look involving doctors into the interpretability of each of these models. Foremost, motivated by the fact that one patient can have multiple pathologies at the same time, a multilabel classification was performed - a task that is not commonly performed by similar studies. The motivation behind considering deep learning and not interpretable non-deep learning techniques is owing to the fact that in recent times deep learning techniques have been observed to outperform others for various radiological applications~\cite{liu2019accurate,yoo2019prostate, dat2019ensembled}.   

The remainder of the paper is organised as follows: in the second section, several related works are presented and discussed, followed by the third section, which details the various network models and interpretability techniques used here and and the approach to dataset creation is delineated. The fourth section presents the classification results and the interpretability analysis. The results are then analysed in the fifth section, and finally, the sixth section concludes the work and provides directions for further research.

\section*{Related works}
The use of artificial intelligence (AI) in healthcare has been developed to support humans in decision making~\cite{vial2018role,davenport2019potential,sloane2020artificial,mahadevaiah2020artificial}. AI-based knowledge has been combined with medical imaging to enhance the accuracy of diagnoses of various diseases, such as respiratory infectious diseases~\cite{agrebi2020use}, pulmonary tuberculosis~\cite{sweetlin2019computer}, including pandemic diseases such as H1N1 influenza~\cite{yao2011computer}. 

The spread of COVID-19 has attracted many researchers to concentrate their efforts toward developing AI-based disease detection techniques for various medical imaging modalities. The assistance of deep learning has shown an improvement in binary diagnosis (presence or absence of COVID-19) from CXR images~\cite{li2020using} and a reduction in the workload of front-line radiologists~\cite{chen2020deep}. Many efforts have been made to perform multiclass classification (COVID-19, other types of pneumonia, or healthy) to assist radiologists in decision making. Narin et al.~\cite{narin2020automatic} used ResNet50, InceptionV3, and InceptionResNetV2 models to classify patients with COVID-19 using CXR images. They demonstrated that the pre-trained ResNet50 model yields the highest accuracy (98\%). However, accuracy is often deemed a misleading metric in the case of imbalanced datasets. Furthermore, they only discriminated between healthy subjects and COVID-19, but did not include the other types of pneumonia. Wang et al.~\cite{wang2020covid} designed COVID-Net using CXR images for the classification of patients with bacterial pneumonia, viral pneumonia, COVID-19, and also healthy subjects with a sensitivity of detection of 91\% COVID-19. Zhang et al.~\cite{zhang2020covid} used a ResNet-based model to classify COVID-19 and non-COVID-19 patients. They achieved a sensitivity of 96\% and a specificity of 70.7\%. Ghoshal et al.~\cite{ghoshal2020estimating} presented a Dropweight-based Bayesian Convolutional Neural Network (BCNN) for CXR-based COVID-19 diagnosis. They found a drastic correlation between the accuracy of the prediction and the uncertainty of the model. Awareness of diagnosis decision uncertainty could endorse deep learning-based applications to be used more and more in clinical routine. Singh et al.~\cite{singh2021interpretable} proposed the Gen-ProtoPNet architecture that provides interpretable classifications of COVID-19 in CXR~\cite{singh2021interpretable} and CT scans~\cite{singh2021object}, resulting in F1 scores as high as 98\%. Furthermore, Shorten et al.~\cite{shorten2021deep} provided a comprehensive survey of different applications of deep learning for COVID-19. On the other hand, De Falco et al.~\cite{de2023classification} proposed an interpretable completely-transparent evolutionary rule-based approach, but only managed to achieve an accuracy of around 80\%. This demonstrates the possible trade-off between transparency and model performance. Deep learning methods that are interpretable, or they are interpreted using post hoc methods, can mitigate this trade-off. Although the application of deep learning methods for COVID-19 lesion detection is not an unexplored topic, including interpretability, systematic comparisons of different models in terms of interpretability and verification of the interpretability results by medical professionals are still missing. these are the aspects this paper seeks to address, while presenting the importance of evaluating or comparing models with respect to interpretability along with the classification accuracy. It is noteworthy that these problems and the message of this paper are not limited to COVID-19 classification, but they are applicable to classification problems in general, especially in high-risk domains like medical imaging.

Although AI-based assistance has been introduced in the field of radiology for a long time, the decision-making mechanisms within these "black-box" methods remains questionable. Recently, research on interpretability has gained more focus. Different interpretability techniques, such as occlusion~\cite{zeiler2014visualizing}, saliency~\cite{simonyan2013deep}, guided backpropagation~\cite{mahendran2016salient}, integrated gradients~\cite{sundararajan2017axiomatic}, etc., have been introduced, demonstrating the potential to open these black boxes.

\section*{Materials and methods}

\subsection*{Network models}
During the course of this research, various network architectures were explored and experimented with, including several variants of VGG~\cite{Simonyan15}, ResNet~\cite{he2016deep}, ResNeXt~\cite{xie2017aggregated}, WideResNet~\cite{zagoruyko2016wide}, Inception~\cite{szegedy2015going}, DenseNet~\cite{huang2017densely}. Prior to training on the dataset of this research work, all the networks were initialised with weights pre-trained on ImageNet. After observing the results, five network architectures were shortlisted for further analysis and also used to create an ensemble using the majority voting strategy for better prediction performance. The models were selected based on different criteria, such as performance, complexity of the model, etc. The selected models are discussed in this section, and Table~\ref{tab:params} shows the complexity of the models.    
\paragraph*{ResNet:}
At the nascent stage of deep learning, the deeper networks faced the problem of vanishing gradients/ exploding gradients~\cite{bengio1994learning,hinton2012improving}, which hampered convergence. The deeper network faced another obstacle called degradation, where the accuracy starts to saturate and degrade rapidly after a certain depth of the network. To overcome these problems, He et al.~\cite{he2016deep} designed a new network model called residual network or ResNet, where the authors came up with `Skip Connection' identity mapping. This does not involve adding an extra hyperparameter or learnable parameter but just adding the output from a preceding layer to a subsequent layer. It unleashed the possibility of training deeper models whilst avoiding these aforementioned issues.

After comparing various versions of ResNet, during this research two different variants, ResNet18 and ResNet34, were chosen for further analysis.
\paragraph*{InceptionNet:}
An image can have thousands of salient features. In different images, the focused features can be in any different part of the image, determining the appropriate kernel size for a convolution network a very difficult task. A large kernel will have a greater focus on globally distributed information, while a smaller kernel will focus on local information.
To overcome this problem, Szegedy et al.~\cite{szegedy2015going} came up with a new network architecture called InceptionNet or GoogleNet. The authors used filters of multiple sizes to operate on the same level, which makes the network more "wider" rather than "deeper".
In order to enhance computational cost-effectiveness, the authors restricted the number of input channels by adding an extra 1x1 convolution before the 3x3 and 5x5 convolutions. Adding 1x1 convolutions is much cheaper than adding 5x5 convolutions.
The authors introduced two auxiliary classifiers to avoid the problem of vanishing gradient, and an auxiliary loss is calculated on each of them. The total loss function is a weighted sum of the auxiliary loss and the real loss.

Excessive reduction in dimensions can cause a loss of information, also known as a "representational bottleneck". To overcome this problem and scale the network in ways that utilise the added computation as efficiently as possible, the authors of InceptionNet introduced a new idea in another publication by Szegedy et al.~\cite{szegedy2016rethinking} factorising convolutions and aggressive regularisation. The authors factored each 5x5 convolution into two 3x3 convolution operations to improve computational speed. Furthermore, they factorised the convolutions of the filter size nxn into a combination of the 1xn and nx1 convolutions. This network is known as InceptionV2.

Szegedy et al.~\cite{szegedy2016rethinking} have also proposed InceptionV3, which extends InceptionV2 further by factorising 7x7 convolutions, label smoothing, and by adding BatchNorm in the auxiliary classifiers. Label smoothing is a type of regularising component added to the loss formula that prevents the network from becoming too confident about a class.

InceptionV3 ranked in one of the top five positions during the initial trials and therefore was used for further analysis.

\paragraph*{InceptionResNetV2:}
The different variants of InceptionNet and ResNet have shown very good performance with relatively low computational costs. With the hypothesis that residual connections would cause Inception network training to accelerate significantly, the authors of the original InceptionNet proposed InceptionResNet~\cite{szegedy2017inception}. In this, the pooling operation inside the main inception modules was replaced by the residual connections. Each Inception block is followed by a filter expansion layer (1x1 convolution without activation), which is used for scaling up the dimensions of the filters back before the residual addition, to match the input size.

This is one of the networks that has been used in this research, because of its performance on the dataset that has been used.
\paragraph*{DenseNet:} Huang et al.~\cite{huang2017densely} came up with a very simple architecture to ensure maximum information flow between layers of the network. By matching feature map size throughout the network, they connected all the layers directly to all of their subsequent layers - a densely connected neural network, or simply known as DenseNet. DenseNet improved the information flow between layers by proposing this different connectivity pattern. Unlike many other networks such as ResNet, DenseNets do not sum the output feature maps of the layer with the incoming feature maps but concatenate them.

In the preliminary trials of this study, DenseNet161 came out as a winner in terms of performance. Therefore, in this research DenseNet161 was included.


\begin{table}[ht]
\caption{\label{tab:params}Number of trainable parameters in each model}
\centering
\begin{tabular}{P{0.20\linewidth}P{0.15\linewidth}P{0.15\linewidth}P{0.15\linewidth}P{0.15\linewidth}}
\toprule
\textit{\textbf{Model}}    & \textit{\textbf{No of parameters}} & \textit{\textbf{GFLOPs}}  & \textit{\textbf{MACs ($\times10^9$)}}  & \textit{\textbf{GPU Memory (Forward + Backward) in GB}} \\ \midrule
\textbf{ResNet18}          & 11,183,694 & 18.95 & 9.53 & 0.15 \\
\textbf{ResNet34}          & 21,291,854 & 38.28 & 19.22 & 0.22 \\
\textbf{InceptionV3}       & 24,382,716 & 35.04 & 17.63 & 0.44 \\
\textbf{DenseNet161}       & 26,502,926 & 80.73 & 40.98 & 1.31 \\
\textbf{InceptionResNetV2} & 54,327,982 & 81.07 & 40.70 & 0.72 \\ \bottomrule
\end{tabular}%
\end{table}
\subsection*{Interpretability techniques}
interpretability techniques can aid in understanding the reasoning of a network for its predictions. In general, the results of interpretability can be visualised using heatmaps, where higher values indicate a heightened focus. However, this may vary among different interpretability techniques. Typically, the heatmaps are overlaid on top of an input image to understand at which parts of the image the network focused to generate the predictions. The techniques that use a single image at a time for analysis are known as local interpretability techniques.
On the other hand, a global interpretability technique often pertains to comprehending how the model works - an aggregated behaviour of the model based on the distribution of the data~\cite{molnar2022interpretable,kopitar2019local}. There are several techniques already in existence. Some of the methods, such as, Occlusion, Saliency, Input X Gradient, Integrated Gradients, Guided Backpropagation, DeepLIFT, Neuron Activation Profiles, which were explored in this research, are explained briefly in this section. 
\paragraph*{Occlusion:}
Occlusion is one of the simplest interpretability techniques for image classifications. This technique helps to understand which features of the image steer the network towards a particular prediction or which are the most important parts for the network to classify a certain image. To obtain this answer, Zeiler et al.~\cite{zeiler2014visualizing} performed an occlusion technique by systematically blocking different parts of the input image with a grey square box and monitoring the output of the classifier. The grey square is applied to the image in a sliding window manner that moves across the image, obtaining many images, and subsequently fed into the trained network to obtain probability scores for a given class for each mask position.

\paragraph*{Saliency:}
In the context of visualisation, saliency refers to a topological representation of the unique features of an image. Saliency is one of the baseline approaches for the interpretation of deep learning models. The saliency method of Simonyan et al.~\cite{simonyan2013deep} returns the gradients of a model for its respective inputs. Positive values present in the gradients show how a small change in the input image changes the prediction.

\paragraph*{Input X Gradient:}
Input X Gradient is an extension of the Saliency approach. Similarly to the saliency method of Simonyan et al.~\cite{simonyan2013deep}, this method of Kindermans et al.~\cite{kindermans2016investigating} also takes the gradients of the output with respect to the input, but additionally multiplies the gradients by the input feature values.
\paragraph*{Guided Backpropagation:}
Guided Backpropagation, also known as guided saliency, is another visualisation technique for deep learning classifiers. Guided backpropagation is a combination of vanilla backpropagation and deconvolution networks (DeConvNet)~\cite{mahendran2016salient}.
In this method, only positive error signals are backpropagated, and the negative signals are set to zero while backpropagating through a ReLU unit~\cite{springenberg2014striving}.
\paragraph*{Integrated Gradients:} Sundararajan et al.~\cite{sundararajan2017axiomatic} proposed a model interpretability technique, which assigns an importance score to each of the features of the input by approximating the integral of the gradients of the output for that input, along the path from the given references for the input.
\paragraph*{DeepLIFT:}
Deep Learning Important FeaTures or DeepLIFT, proposed by Shrikumar et al.~\cite{shrikumar2017learning}, is a method to pixel-wise decompose the output prediction of a neural network on a specific input. This involves backpropagating the contributions of all neurons in the network to every feature of the input. DeepLIFT compares the activation of each neuron to its "reference activation", and then assigns contribution scores based on the difference. DeepLIFT can also reveal dependencies that might be missed by other approaches by optionally assigning separate considerations to positive and negative contributions. Unlike other gradient-based methods, it uses difference from reference, which permits DeepLIFT to propagate an importance signal even in situations where the gradient is set to zero.

\paragraph{Neuron Activation Profiles:}

The aforementioned interpretability techniques are local methods that help to understand single predictions of a neural network.
To investigate model behaviour more generally, a global interpretability technique called Neuron Activation Profiles (NAPs) is employed~\cite{Krug2018b, krug2021snaps}.
NAPs describe and contrast the activity of the neural network of sets of related inputs, for example, of different classes, using an averaging approach.
Initially, the activation values in the layers of interest are obtained by computing a forward pass for every test image.
Then, the average feature maps over each respective group are computed to characterise the group-specific activity. 
In addition to characterising the network activations for a group, further emphasis is given to the differences between the groups.
To this end, the average over all groups is subtracted from each group's average.
These normalised averaged activation values can be interpreted as the activation difference from the global average.
Positive values indicate a characteristically high neuron activation compared to the entire data set, and negative values indicate a comparably low neuron activation.
NAP values are particularly useful to identify which activations differ between groups of interest and correspondingly indicate the model's ability to distinguish between the classes according to the activations.
When working with image data, visually interpretable plots of NAPs of feature maps can be created. 
For data that are not visually interpretable, NAPs can be further used for similarity analyses~\cite{krug2021snaps} or for dimensionality reduction-based visualisation~\cite{krug2022visualizing}.

In order to obtain useful averaging results, this method requires data in which the objects are at the same location in the images.
This alignment is guaranteed through data preprocessing that resizes and crops the original images.


\subsection*{Implementation}
The models were implemented using PyTorch~\cite{NEURIPS2019_9015}. An interpretability pipeline for PyTorch-based classification models was developed with the help of Captum~\cite{kokhlikyan2020captum}. The code of this project is available on GitHub:~\url{https://github.com/soumickmj/diagnoPP}. The pipeline was later made part of the TorchEsegeta~\cite{chatterjee2021torchesegeta}. 

Training sessions were conducted using Nvidia GeForce 1080 Ti and 2080 Ti GPUs, each with 11GB of memory. The loss was calculated using Binary Cross-Entropy (BCE) with Logits, which combines the sigmoid layer with the BCE loss, to achieve better numerical stability than using the Sigmoid layer followed by BCE loss separately. The numerical stability is achieved by using the log-sum-exp trick, which can prevent underflow/overflow errors. The loss was minimised by optimising the model parameters using the Adam optimiser~\cite{kingma2014adam}, with a learning rate of 0.001 and a weight decay of 0.0001. A manual seed was used to ensure the reproducibility~\cite{pytorchrepro} of the models. Automatic Mixed Precision was used using Apex~\cite{apex}, to speed up training and decrease GPU memory requirements. 

The interpretability methods were applied on the models using Nvidia Tesla V100 GPUs, having 32GB memory each. Some of the interpretability techniques could not be used on certain models owing to insufficient GPU memory caused by the complexities of the models.

\subsection*{Data}
\subsubsection*{Data Collection}
\begin{figure}[!t] 
\centering
\includegraphics[width=0.6\textwidth]{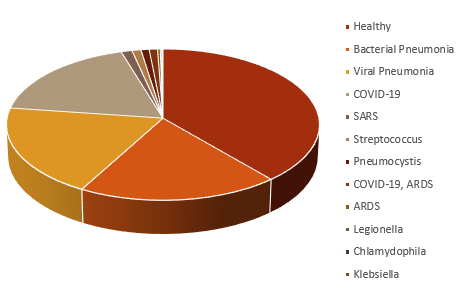}
\caption{CXR images distribution for each infection type in the dataset}
\label{fig:dataset}
\end{figure}

The CXR images were collected from two public datasets. The first dataset was the COVID-19 image data collection by Cohen et al.~\cite{cohen2020covid,covidDS}, comprising 236 images of COVID-19, 12 images of COVID-19 and ARDS, 4 images of ARDS, 1 image of Chlamydophila, 1 image of Klebsiella, 2 images of Legionella, 12 images of Pneumocystis, 16 images of SARS, 13 images of Streptococcus, and 5 images without any pathological findings. The second dataset was the Chest X-ray Images (Pneumonia) dataset by Kermany et al.~\cite{kermany2018labeled,pneumoniaDS}, which has a total of 1583 images of healthy subjects, 1493 images of viral pneumonia and 2780 of bacterial pneumonia. From this dataset, 500 images of healthy, 250 images of viral pneumonia, and 250 images of bacterial pneumonia were randomly chosen. Fig~\ref{fig:dataset} portrays the final data distribution considered for the work. This CXR image dataset comprises posterior anterior (PA), anterior superior (AP), and anterior superior supine (AP supine) radiographs. Whilst the AP view is not the preferred positioning and has disadvantages such as organ overlap that could interfere with network prediction~\cite{cxr}, it is a technique commonly used for COVID-19 patients in a coma. 

\begin{figure}[!t]
\centering
\includegraphics[width=0.7\textwidth]{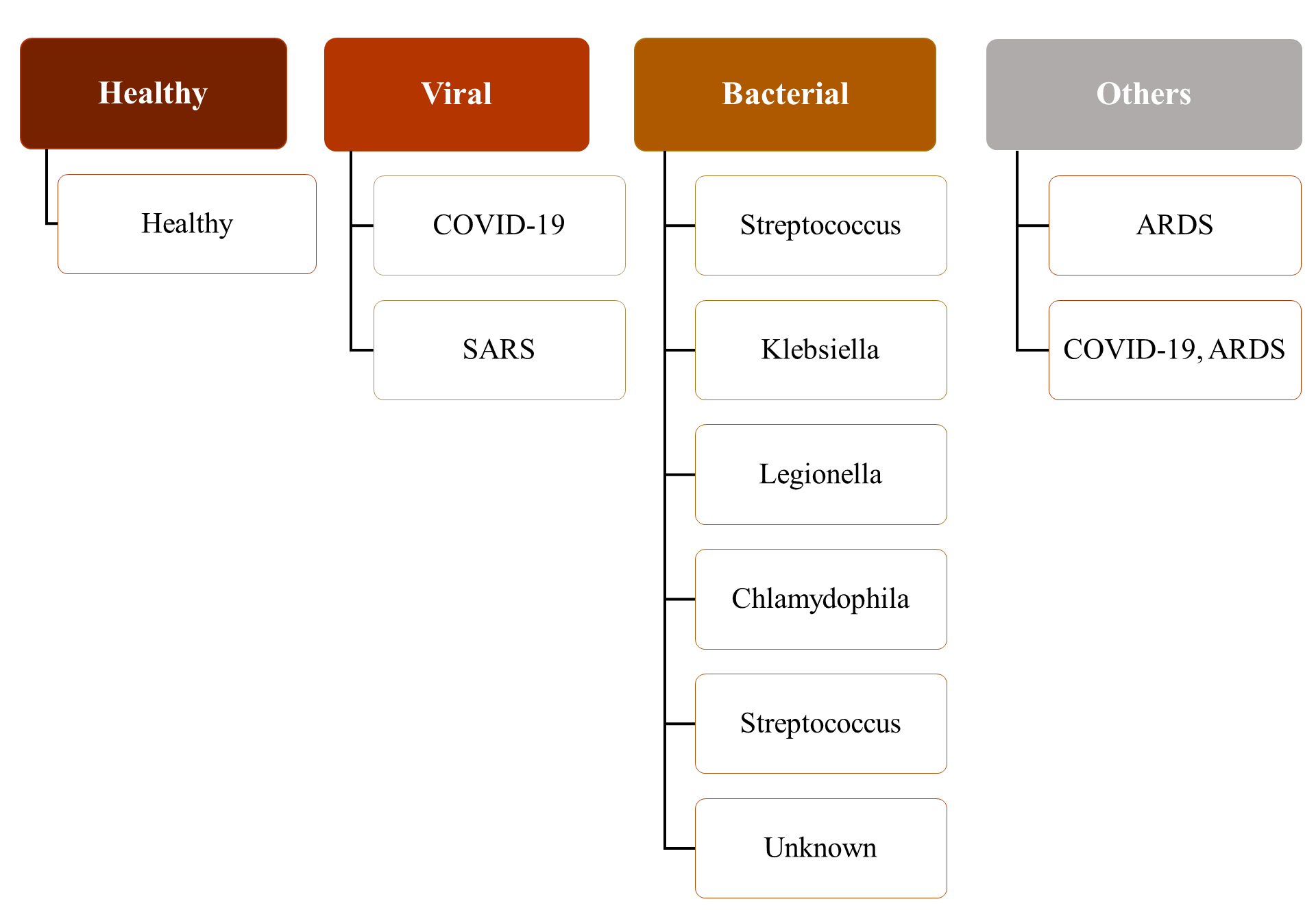}
\caption{A hierarchy of pathological labels used in this study}
\label{fig:labelsHierarchy}
\end{figure}

The hierarchical nature of the pathologies can be observed in this combined dataset (see Fig~{\ref{fig:labelsHierarchy}}). For example, SARS and COVID-19 are subtypes of viral pneumonia. However, Streptococcus, Klebsiella, Chlamydophila, and Legionella are subtypes of bacterial pneumonia, and Pneumocystis is a subtype of fungal pneumonia. Furthermore, viral, bacterial, and fungal pneumoni{\ae} are different types of pneumonia. Therefore, a patient having COVID-19 inherently has viral pneumonia. ARDS, which stands for acute respiratory distress syndrome, is a serious lung condition with a high mortality rate~\cite{diamond2021acute}. It frequently develops alongside pathological conditions like nonpulmonary sepsis, aspiration, or pneumonia~\cite{matthay2019acute}. Although the respiratory pathologies of ARDS (associated with or without COVID-19) and COVID-19 are similar, COVID-19 has different features that require different patient management, and a patient suffering from both could require additional care~\cite{fan2020covid,gattinoni2020covid,bain2021covid}. Therefore, the dataset, which comprises cases where a patient has both COVID-19 and ARDS, is suitable for multilabel classification.

\subsubsection*{Dataset Preparation}
The final dataset was randomly divided into a training set, consisting of 60\% of unique subjects, and the remaining 40\% of the subjects were used as a test set. Five-fold cross-validation (CV) was conducted to assess the generalisation capabilities of the models. The performance of the models during the 5-fold CV is reported in the sub-section \ref{ModelOutcome}. For the interpretability analysis, only the results from the first fold were used, as this yielded the highest micro F1 scores.

\subsubsection*{Pre-processing}

The dataset used for the task comprises X-ray images collected at different centres using different protocols and varying in size and intensity. Therefore, all the images were initially pre-processed to have the same size. To make the image size uniform throughout the dataset, each image was interpolated employing bicubic interpolation, to have 512 pixels on the longer side. the pixel count on the shorter side was determined, keeping the aspect ratio of the original image. Subsequently, zero-padding was applied to the shorter side to make that side have 512 pixels, resulting in a 512 x 512 image. Image resizing was followed by percentile cropping, where the image intensity was cropped to the first and 95th percentile, and then the intensity normalisation was performed to the range [0,1]. The percentile cropping normalisation minimises the effect of intensity variation due to non-biological factors.

\subsubsection*{Classification Setup}
In this multilabel classification setup, the model was trained to identify the disease and also its supertypes. Therefore, when a network encounters an image of a COVID-19 patient, it should ideally predict it as pneumonia, viral pneumonia, and COVID-19. When a network encounters an image of a patient having multiple pathologies, as in this dataset, some patients have both COVID-19 and ARDS, ideally, the network should classify it as pneumonia, viral pneumonia, COVID-19, as well as ARDS. Interpretability analysis was conducted for each label of each image in the test set. 

\subsection*{Evaluation Metrics}
In a multiclass setting, classifiers are generally evaluated with respect to precision, recall and F1 metrics. In a multilabel classification setting, these metrics are computed in two manners: macro and micro averaging~\cite{tsoumakas2007multi}. 
\begin{equation}
    Macro = \frac{1}{P}\sum_{i=1}^{p}Metric\Bigg(TP_i, FP_i, TN_i, FN_i\Bigg)
    \label{eqn:macro}
\end{equation}
As shown in Eq.~\ref{eqn:macro}, the macro-based metrics are first computed individually from the true positives (TP), true negatives (TN), false positives (FP) and false negatives (FN) of each class/pathology and then averaged, where \textit{P} denotes the number of classes and \textit{Metric} $\in$ \{precision, recall, F1\}.

This manner of computation of metrics helps to treat each pathology equally, and the metric values are significantly influenced by the rarer labels.
 \begin{equation}
Micro = Metric(\sum_{i=1}^{p}TP_i, \sum_{i=1}^{p}FP_i, \sum_{i=1}^{p}TN_i, \sum_{i=1}^{p}FN_i)
\label{eqn:micro}
\end{equation}
In micro-based metrics, TP, TN, FP, and FN of each class/pathology are added individually and then averaged, as shown in Eq.~\ref{eqn:micro}. Therefore, the micro-based metrics portray the aggregated contribution of all classes/pathologies. Therefore, the influence of the predictions from the minority classes becomes diluted among the contributions from the majority classes. This makes the micro-based metrics a suitable measure for estimating the overall performance of the classifier, particularly in scenarios involving imbalanced datasets. Given the significant imbalance in the utilised dataset, micro-based metrics have been considered for classifier evaluation~\cite{charte2015addressing}. 

\section*{Results}
\subsection*{Model outcome}\label{ModelOutcome}
\subsubsection*{Overall comparisons of the classifiers}

Fig.~\ref{fig:CompClassify} shows that the overall performance of the classifiers over pathologies was similar. Among the non-Ensemble models, DenseNet161 performed the best in all metrics. Although InceptionResNetV2 was the most complex model among all, it yielded the poorest recall, which implies that the ability of the model to find pathology-affected cases was poor compared to less complex models. ResNet18 was the least complex model among the non-Ensemble classifiers, ranking second to DenseNet161 with respect to micro F1.  The ensemble produced the best results and the minimum variance as presented in Table \ref{tab:micro} in the 5-fold cross-validation.

\begin{table}[ht]
\caption{Performance of all the classifiers with respect to micro based metrics\label{tab:micro} over 5-folds}
\centering
\begin{tabular}{P{0.20\linewidth}P{0.20\linewidth}P{0.20\linewidth}P{0.20\linewidth}}
\toprule
\textit{\textbf{Model}}    & \textit{\textbf{Precision}} & \textit{\textbf{Recall}} & \textit{\textbf{F1}} \\ \midrule
\textbf{DenseNet161}       & 0.864 $\pm$ 0.012                   & 0.845 $\pm$ 0.015                & 0.854 $\pm$ 0.008           \\
\textbf{InceptionResNetV2} & 0.844 $\pm$ 0.023                  & 0.787 $\pm$ 0.063               & 0.814 $\pm$ 0.042           \\
\textbf{InceptionV3}       & 0.802 $\pm$ 0.065                   & 0.792 $\pm$ 0.044               & 0.796 $\pm$ 0.053           \\
\textbf{ResNet18}          & 0.824 $\pm$ 0.014                  & 0.824 $\pm$ 0.008              & 0.824 $\pm$ 0.007           \\
\textbf{ResNet34}          & 0.815 $\pm$ 0.022                  & 0.800 $\pm$ 0.025               & 0.807 $\pm$ 0.018           \\ \midrule
\textbf{Ensemble}          & 0.889 $\pm$ 0.010                  & 0.851 $\pm$ 0.005              & 0.869 $\pm$ 0.007           \\ \bottomrule
\end{tabular}
\end{table}

Another interesting observation that could be made is regarding inactive feature maps (dead neurons). DenseNet161 had the highest percentage of such feature maps - as high as 99.22\% for the middle layer. Although InceptionResNetv2 was the most complex, it had fewer inactive feature maps than DeseNet161. ResNets, the least complex models in this study, had the lowest percentage of inactive feature maps (48.44\% and 60.16\% for the middle layers of ResNet18 and ResNet34, respectively).

\subsubsection*{Comparisons of the classifiers for different pathologies}
\begin{figure}
     \centering
     \begin{subfigure}[b]{0.3\textwidth}
         \centering
         \includegraphics[width=\textwidth]{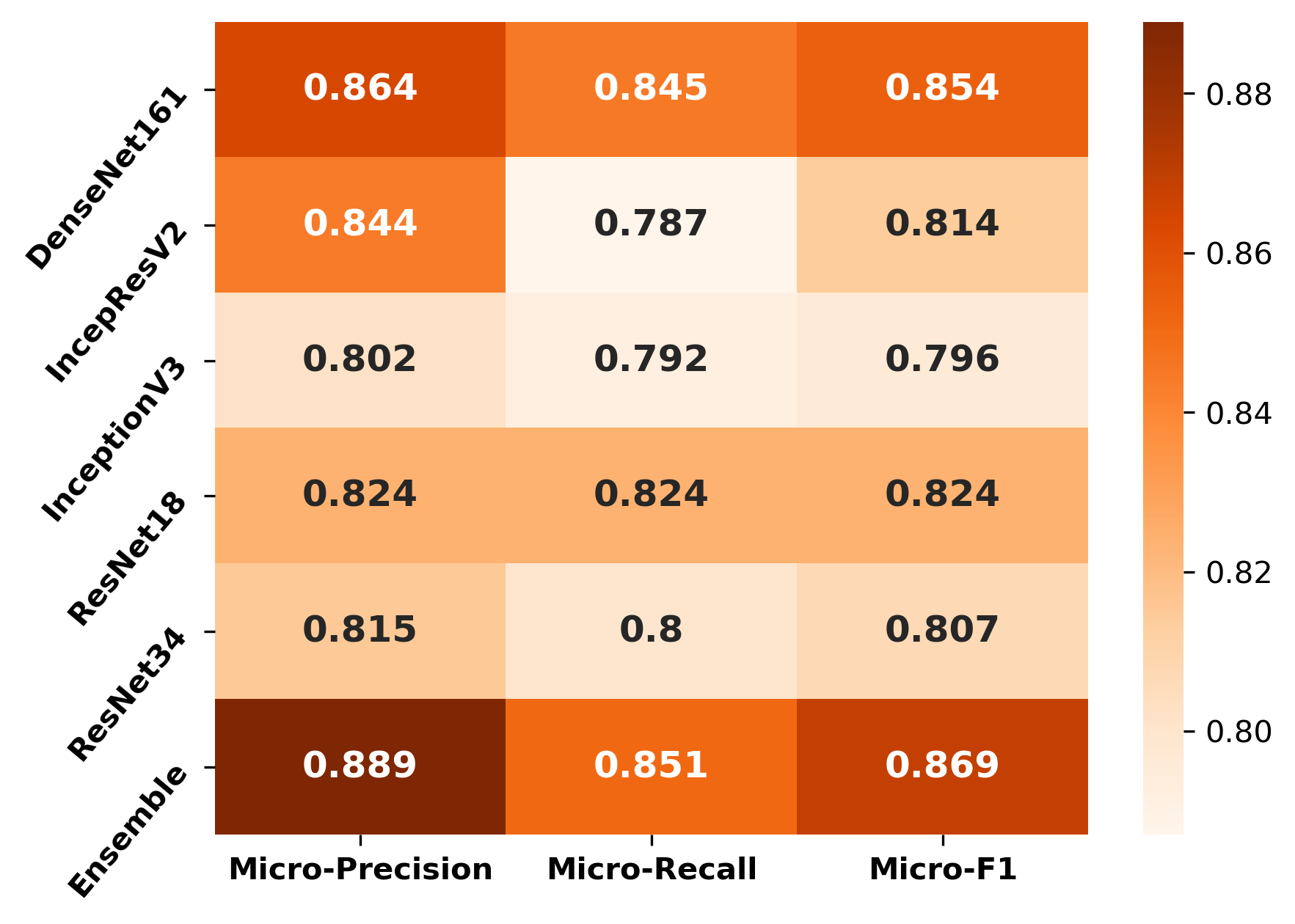}
\caption{Micro metrics}
\label{fig:CompClassify}
     \end{subfigure}
     \hfill
     \begin{subfigure}[b]{0.3\textwidth}
         \centering
         \includegraphics[width=\textwidth]{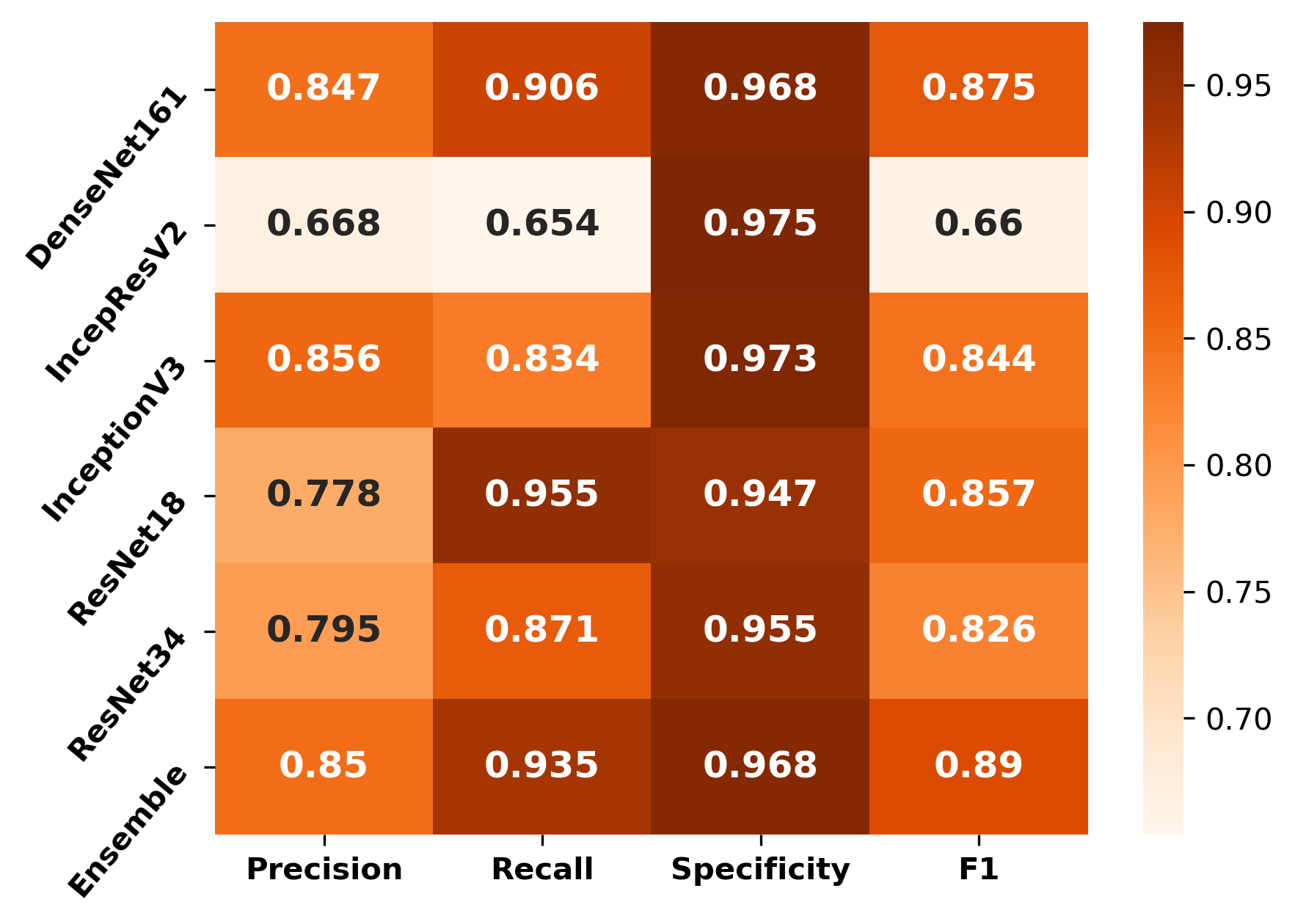}
\caption{COVID-19}
\label{fig:mat_covid}
     \end{subfigure}
     \hfill
     \begin{subfigure}[b]{0.3\textwidth}
         \centering
         \includegraphics[width=\textwidth]{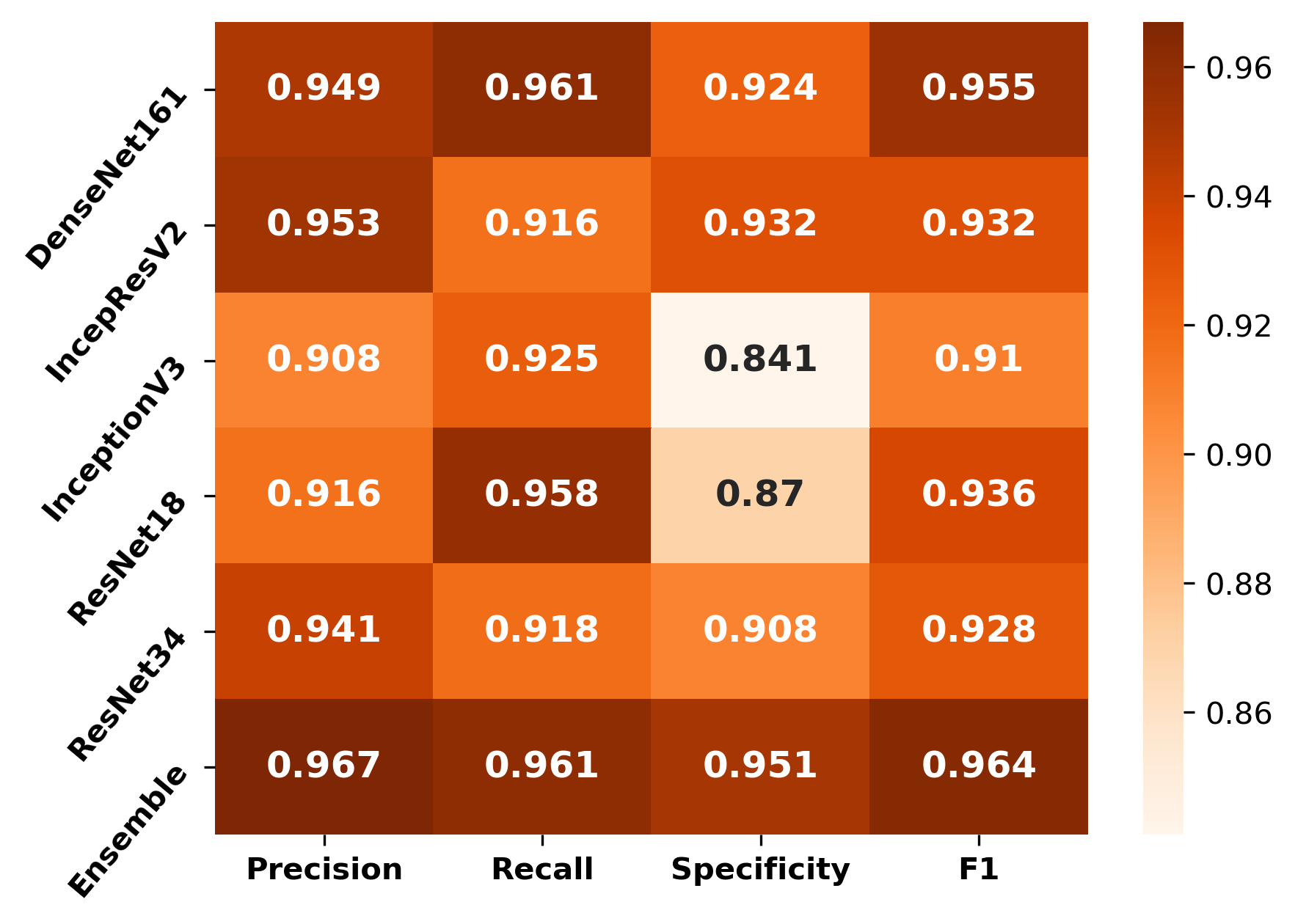}
\caption{Pneumonia}
\label{fig:mat_Pneumonia}
     \end{subfigure}
     \hfill
     \begin{subfigure}[b]{0.3\textwidth}
         \centering
         \includegraphics[width=\textwidth]{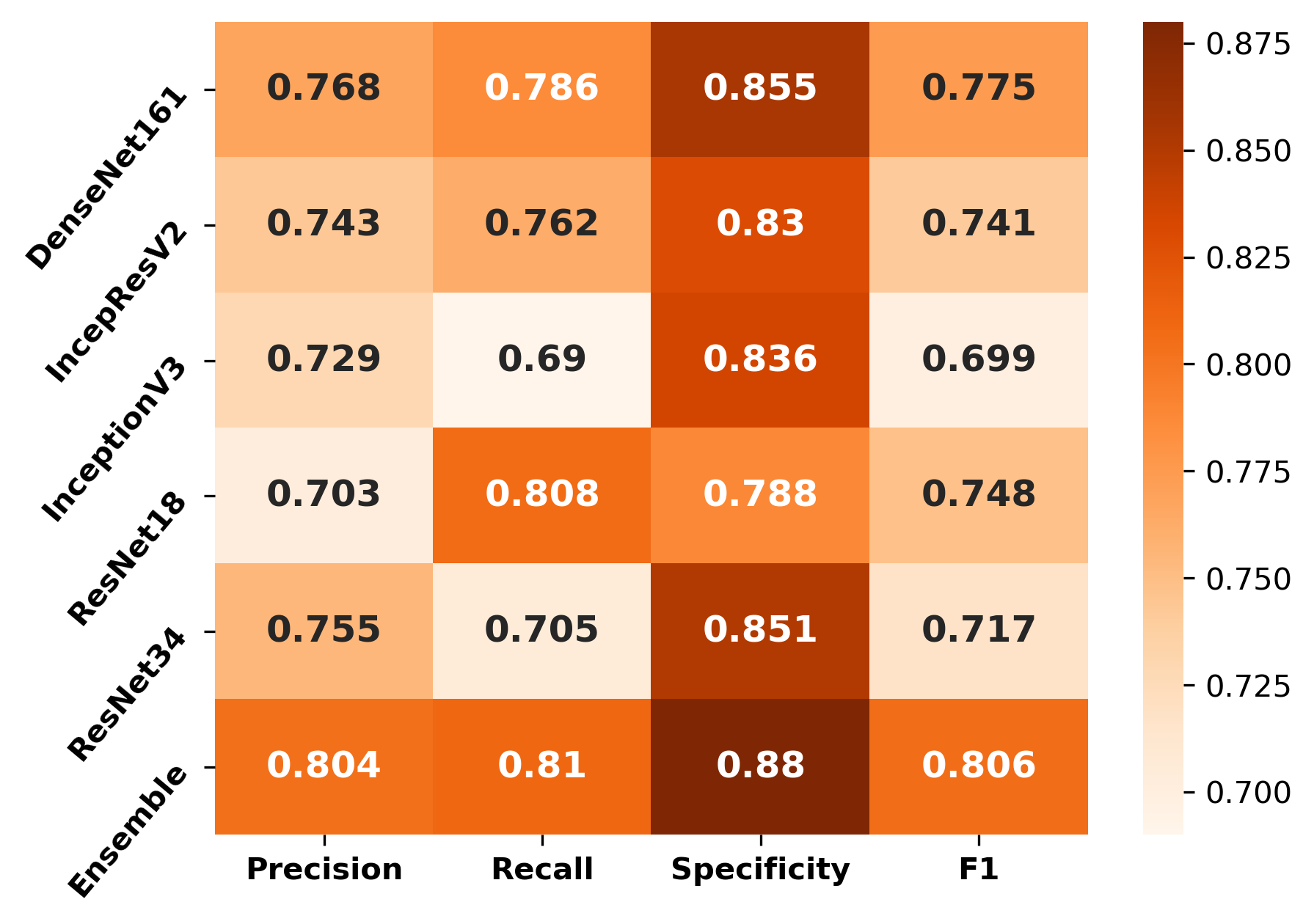}
\caption{Viral Pneumonia}
\label{fig:mat_ViralPneumonia}
     \end{subfigure}
     \hfill
     \begin{subfigure}[b]{0.3\textwidth}
         \centering
         \includegraphics[width=\textwidth]{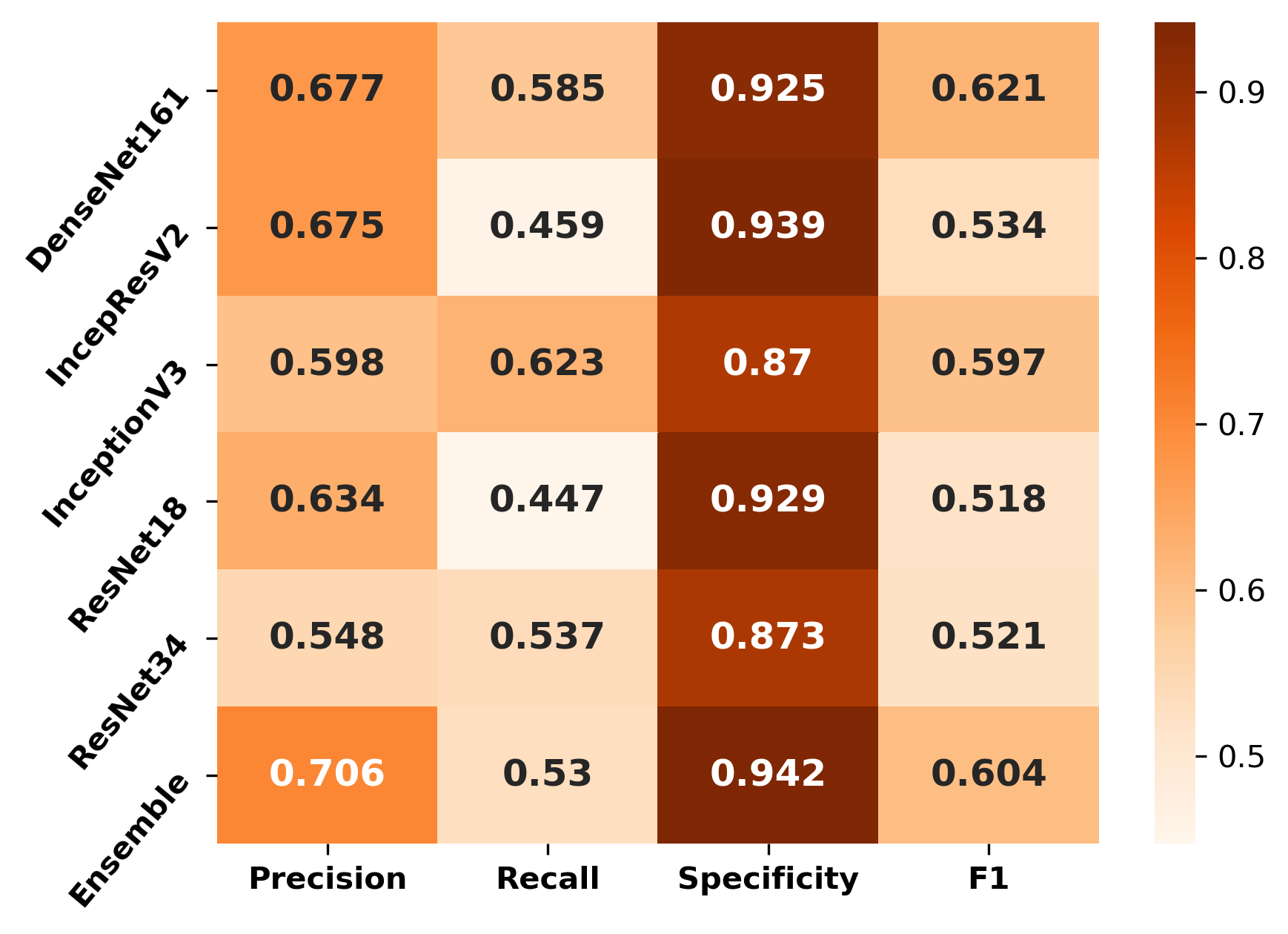}
\caption{Bacterial Pneumonia}
\label{fig:mat_BacterialPneumonia}
     \end{subfigure}
     \hfill
     \begin{subfigure}[b]{0.3\textwidth}
         \centering
         \includegraphics[width=\textwidth]{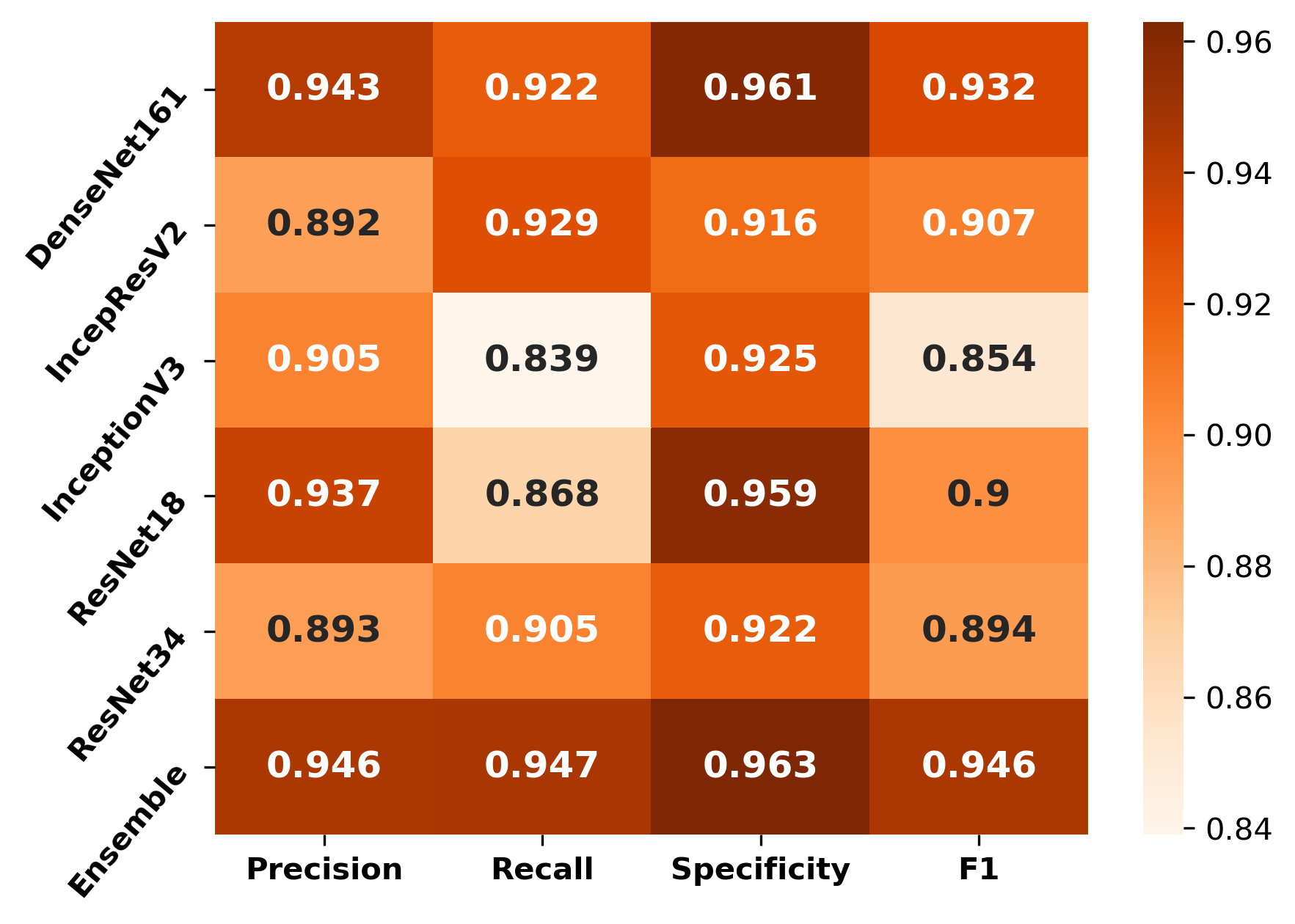}
\caption{Healthy subjects}
\label{fig:mat_Healthy}
     \end{subfigure}
        \caption{Comparison of the classifiers based on micro metrics (a) and their performance for the different classes (b-f)}
        \label{fig:conf_mat}
\end{figure}

The authors also compared the classifiers' performance at the pathology level. The average metric values across five cross-validation folds have been depicted in Fig~\ref{fig:mat_covid} to Fig~\ref{fig:mat_Healthy} for COVID-19, pneumonia, viral pneumonia, bacterial pneumonia, and healthy subjects, respectively. When comparing the models using the average F1, it has been observed that the performance of most models for COVID-19, pneumonia, and healthy was good, except for the performance of InceptionResNetV2 for COVID-19 cases. Among all models, the results of DenseNet161 were the most promising for all diseases. For the COVID-19 classification, DenseNet161 performed the best, and ResNet18 bagged the second position. DenseNet161 performed the best for pneumonia. InceptionResNetV2 provided the highest performance for the classification of viral pneumonia. Lastly, InceptionV3 gave the highest scores for bacterial pneumonia. 

\subsection*{Interpretability of models}
In the first sub-subsection \ref{int1.1} different interpretability techniques have been explored for different classifiers with respect to the different diseases. 
The second subsection \ref{int1.2} talks about how the different models performed for specific pathologies.

All the given interpretability analyses (except using the global method NAP) were performed for that specific input CXR image which has been shown as the underlay. In the interpretability analysis using NAP, all images from the test set were used, as this method performs a global analysis. 

\subsubsection*{Pathology based comparisons of local interpretability techniques for models}
\label{int1.1}
\begin{figure}[!t]
\centering
\includegraphics[width=0.95\textwidth]{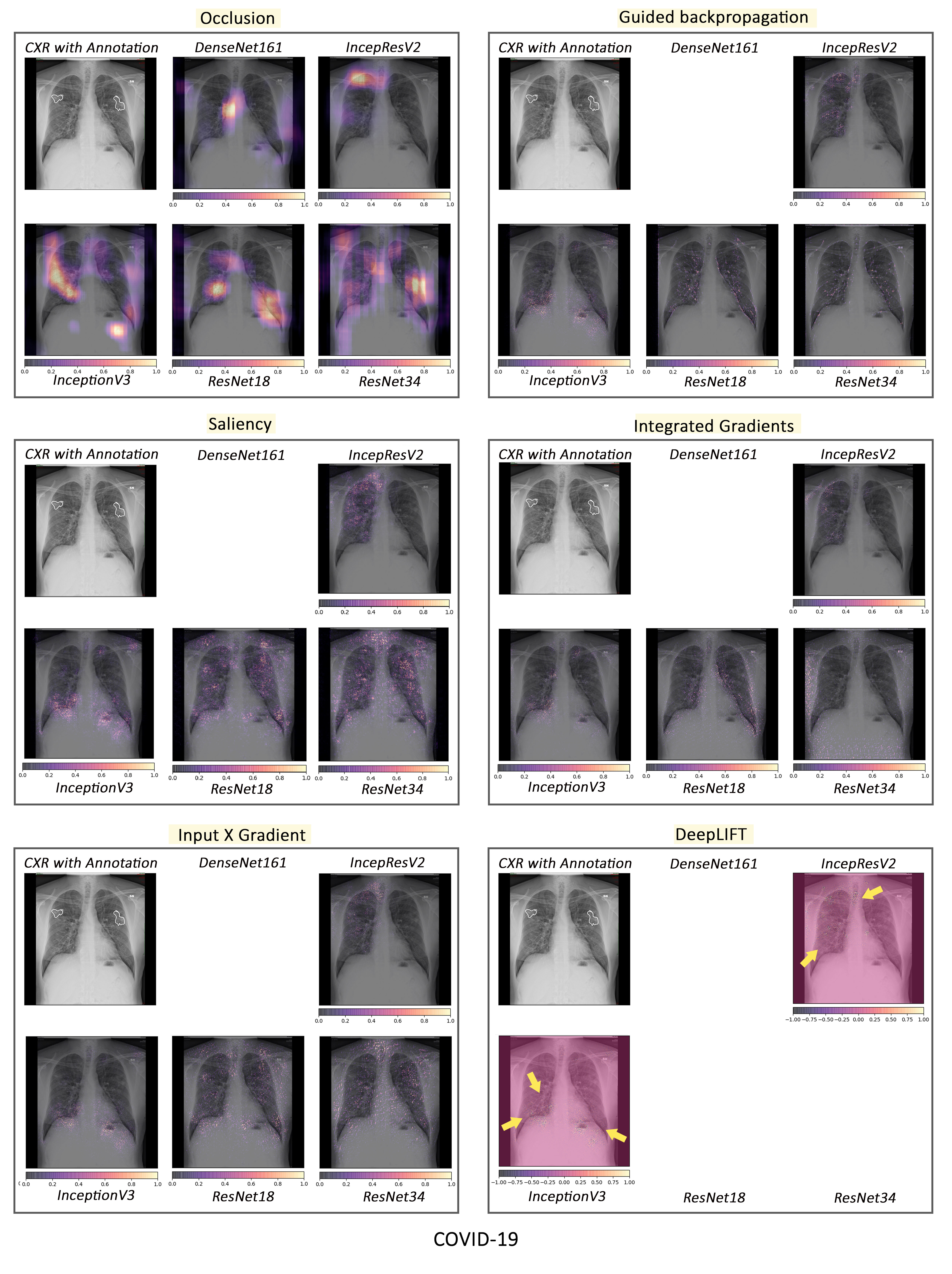}
\caption{Comparison of various interpretability techniques with respect to models for COVID-19 predictions against the manual annotation of the affected areas by medical experts.}
\label{fig:pathoCompareCovidAllTech}
\end{figure}

\begin{figure}[!t]
\centering
\includegraphics[width=0.95\textwidth]{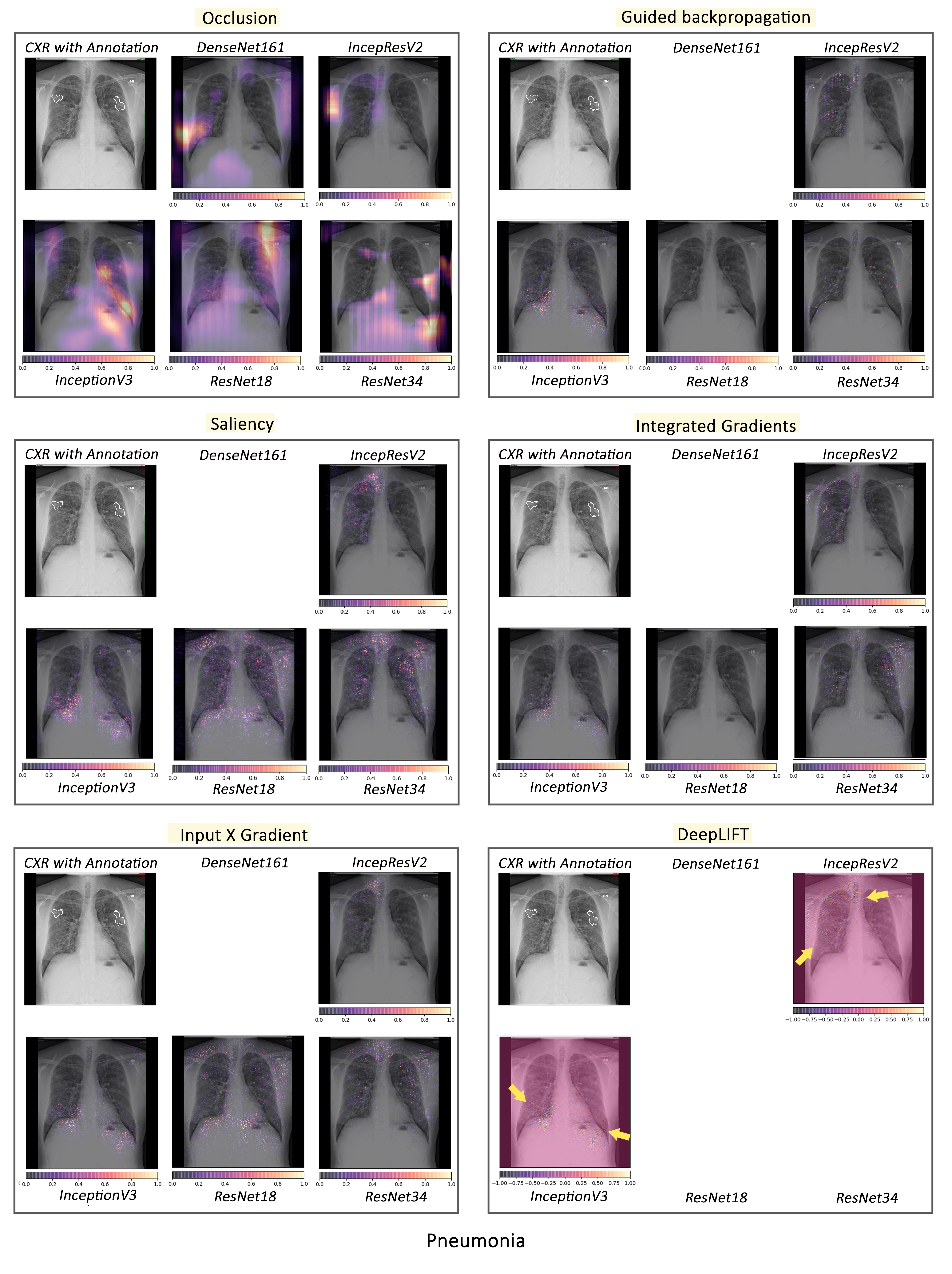}
\caption{Comparison of various interpretability techniques with respect to models for pneumonia predictions against the manual annotation of the affected areas by medical experts.}
\label{fig:pathoComparePneumoAllTech}
\end{figure}

\begin{figure}[!t]
\centering
\includegraphics[width=0.95\textwidth]{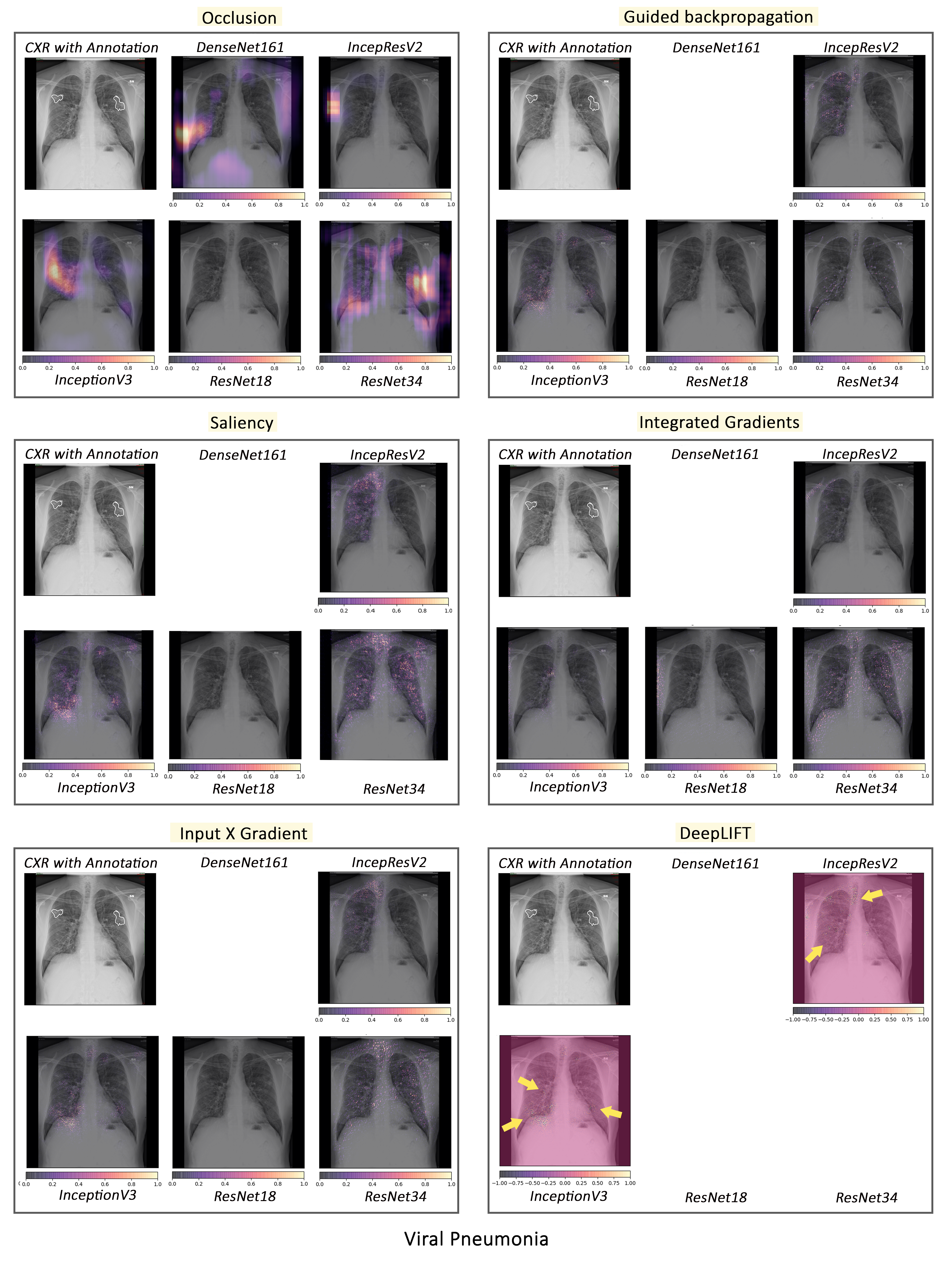}
\caption{Comparison of various interpretability techniques with respect to models for viral pneumonia predictions against the manual annotation of the affected areas by medical experts.}
\label{fig:pathoCompareViralPneumoAllTech}
\end{figure}

To visualise the results for a specific case, the models were interpreted using local methods: occlusion, saliency, inputXgradient, guided backpropagation and integrated gradients, and have been shown in Fig.~\ref{fig:pathoCompareCovidAllTech}, Fig.~\ref{fig:pathoComparePneumoAllTech} and Fig.~\ref{fig:pathoCompareViralPneumoAllTech}. Apart from occlusion, the other interpretability techniques failed to run for DenseNet161 due to GPU memory limitations. in DeepLIFT, ResNets faced an additional challenge due to the ReLU operations used "in place" in those models. Models have to be updated to run DeepLIFT on them.

According to the clinical findings of the COVID-19 image data provided by Cohen et al.~\cite{cohen2020covid}, multiple abnormalities of the lungs were located in the upper and lower pulmonary field, as well as the upper left part of the lung. The models classified this case as COVID-19, pneumonia, and viral pneumonia responding to the pathology of lung infection. It can be seen that the focus area of the models for COVID-19 differs from the focus area for pneumonia and viral pneumonia. DenseNet161 and InceptionResNetV2 focused primarily on the right lung. InceptionV3, ResNet18, and ResNet34 covered both the right and left parts, not only the lesion but also the irrelevant regions outside the lung.

Local interpretability methods suffered mainly from false positives. In some cases, the occlusion did not detect the affected areas for DenseNet161 and InceptionResNetV2 and falsely marked the normal areas as positive, as shown in Fig.~\ref{fig:pathoCompareCovidAllTech}. Furthermore, for InceptionV3, it detected some positive patches, but falsely detected more areas as positive. Finally, in general, for ResNets, occlusion was most sensitive to positive areas and detected lesser false negatives.
Guided backpropagation, saliency, integrated gradients, and DeepLIFT in general falsely detected normal lung areas as positive - picked up normal bronchovascular markings as positive and did not mark the actual affected areas.
The input X gradient detected some positive areas correctly for ResNet18, but falsely marked many normal areas. In general, the representations learnt by the ResNet models captured the most accurate regions as seen from most interpretability techniques, with fewer false negatives. Among the local interpretability techniques, occlusion provided the best guidance in finding clinically important areas, which were confirmed by medical experts. 


\subsubsection*{Intense Interpretability}
\label{int1.2}

\paragraph{The failure case of the best performing model for COVID-19 classification:}

\begin{figure}[!htbp]
\centering
\includegraphics[width=\textwidth]{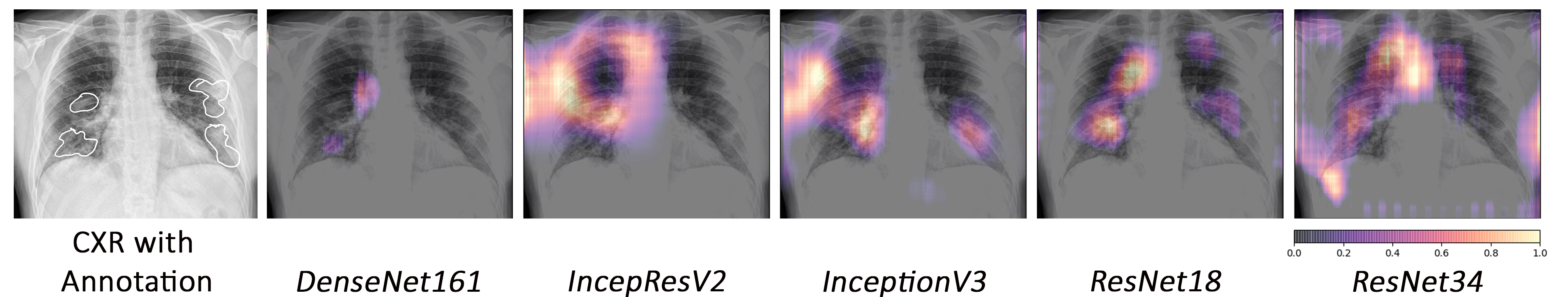}
\caption{A case-study of DenseNet161 failure using occlusion. The affected areas in the lungs have been annotated by medical experts.}
\label{fig:DenseNet161_failureCase}
\end{figure}

Although DenseNet161 performed the best among all models, it gave false negatives for some of the COVID-19 patients, while the rest of the models, including the ensemble, could correctly predict. The occlusion results of the models can be observed in Fig.~\ref{fig:DenseNet161_failureCase}. This figure shows that DenseNet161 and InceptionResnetV2 did not focus on any affected areas, but rather on other regions (e.g. normal right hilum). InceptionV3, ResNet18, and ResNet34 mainly focused on affected areas with good sensitivity. InceptionV3, however, had more false positives than ResNets (e.g. outside the right lung).

Another analysis was performed with CXR of a 70-year-old woman who had three days of cough, myalgia, and fever; without any recent overseas travel. A series of chest radiographs were obtained before confirmation of coronavirus infection, and follow-ups were done in three days, seven days, and nine days. It shows the progression of radiographic changes. In the image prior to COVID-19, both models falsely detected all normal areas as relevant features. In the image of day 3, the doctor could not visually detect any affected area, although this was the image from the third day after testing positive for COVID-19. This might indicate that when no substantial affected area can be seen in the image visually (i.e., day 3), the model might have been picking up some mild markers, which visually cannot be confirmed. In the images of days seven and nine, DesNet161 did not focus correctly on the affected regions and had both false positives and false negatives, while ResNet18 focused on the affected regions more accurately.

\begin{figure}[!h]
\centering
\includegraphics[width=0.95\textwidth]{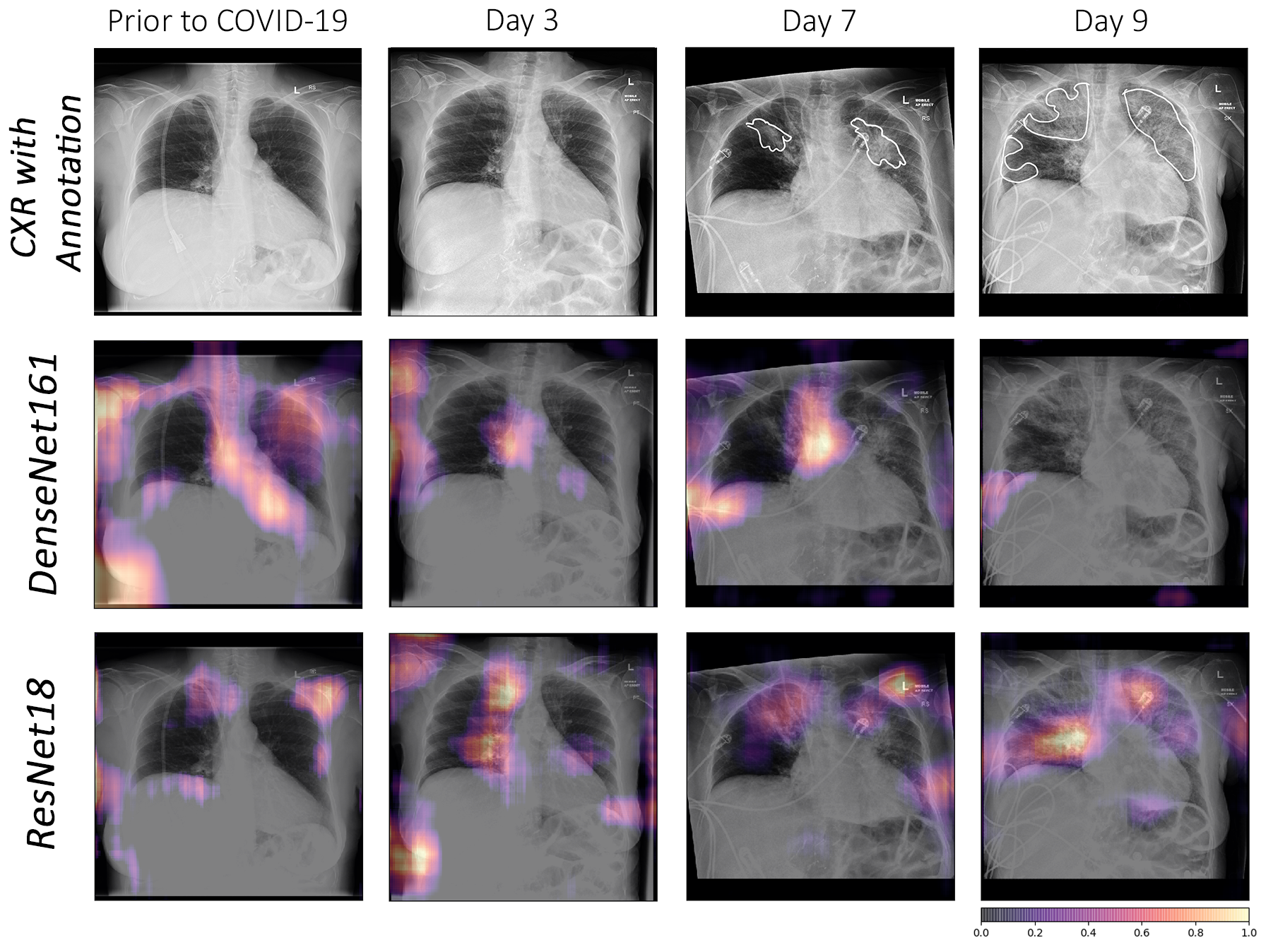}
\caption{Comparison using occlusion between DenseNet161 and ResNet18 for a specific COVID-19 follow-up case. The affected areas in the lungs have been annotated by medical experts.}
\label{fig:FollowUpDenseNet161VSResNet18}
\end{figure}

ResNet18 can be considered the overall winner, as it yielded high evaluation scores, despite having the least number of network parameters. Furthermore, its interpretability analysis showed the location of the lesion, which allows to use this network for follow-up or severity estimations, as illustrated in Fig.~\ref{fig:FollowUpDenseNet161VSResNet18}.  

\paragraph{Representations in DenseNet161 and ResNet18:}
In addition to individual failure cases, the authors further investigated how the COVID-19 and pneumonia pathologies are represented in the neuron activations of DenseNet161 and ResNet18 compared to healthy individuals. This representation analysis was performed using NAPs - a global interpretability technique.
In general, in a well-generalised model, larger neuron activation differences are expected between different pathologies and healthy subjects in the lungs than in other image areas.
If activity differences are observed in other regions, this indicates that the model exploits biologically irrelevant features to discriminate the classes.

\begin{figure}
\centering
\includegraphics[width=\textwidth]{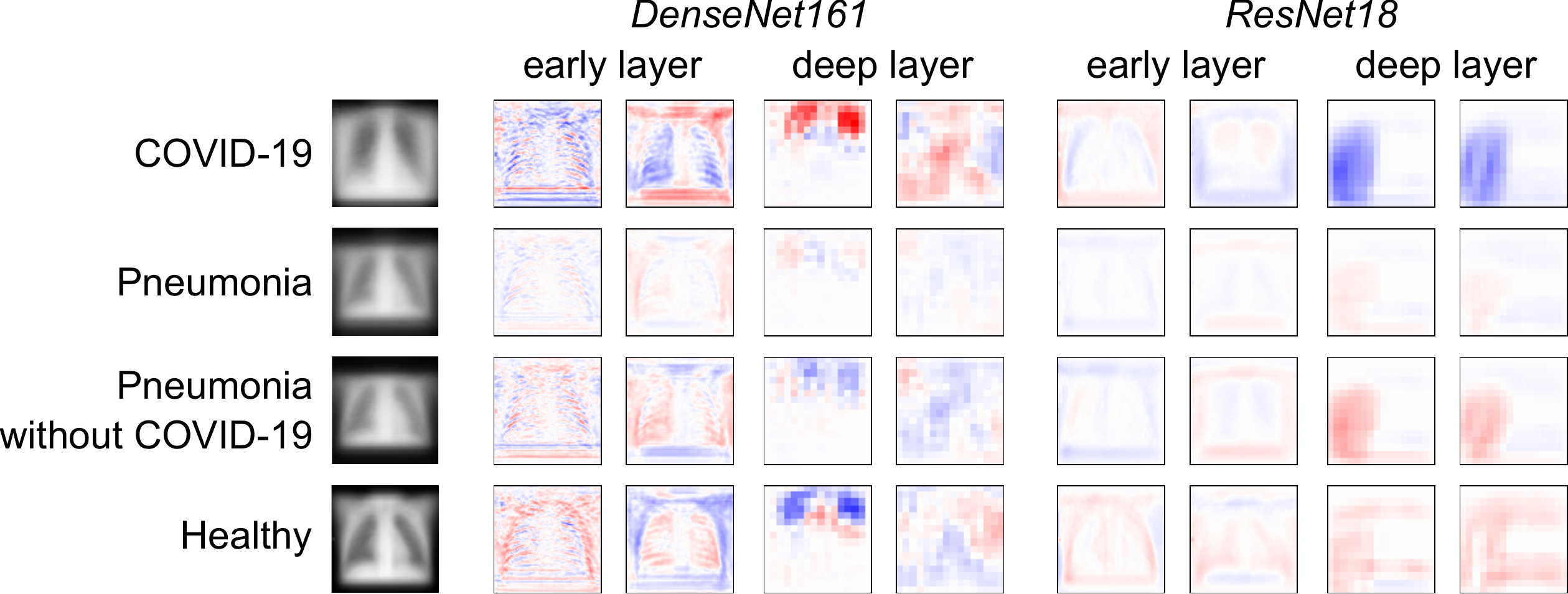}
\caption{Average input images and feature map NAPs in different models and layers for different pathologies and healthy subjects. Blue indicates lower activation of the respective neuron for this group compared to the other groups, red indicates higher activity.}
\label{fig:naps}
\end{figure}

To find potentially exploitable features, the input averages (input layer NAPs) are first investigated in Fig.~\ref{fig:naps} (left). It can be observed that pneumonia images cover a smaller portion of the height dimension than COVID-19 or healthy subjects images. This means that there are dark top and bottom regions in the majority of Pneumonia images. Based on this observation, the authors hypothesised that a model might exploit this non-biological feature.

To investigate this hypothesis, the feature map NAPs of DenseNet161 and ResNet18 in an early and deep layer, respectively, are visualised.
The authors particularly investigate layers at representative depths of the networks.
For DenseNet161, the ReLU-activated outputs of the first and last dense blocks were chosen.
As representative layers of ResNet18, the outputs after the first and last residual connections were selected.
For these layers, two exemplary feature map NAPs among those of the highest activity differences between the observed classes are shown in Fig.~\ref{fig:naps}. In DenseNet161, one can clearly observe activation differences in both the border regions and the lung.
For example, COVID-19 images are easy for the model to distinguish based on the activation difference corresponding to not having dark regions at the bottom and top of the images.
In the deeper layer, the activation difference patterns do not resemble any interpretable structure, neither in the lungs nor in the lower and upper regions.
This indicates why DenseNet161 has a high performance despite giving false negative COVID-19 results.
Instead of detecting COVID-19-specific features, it likely exploits features of the data that are correlated but not related to the pathology.
However, it does not appear that DenseNet161 uses dark border regions as the main distinguishing factor.
ResNet18, in contrast, is less likely to detect biologically irrelevant features.
Although in the early layers there are activation differences in the top and bottom areas of the images, in most deep-layer feature maps, the groups can be most clearly distinguished from each other from neuron activity in the (right) lung regions.

\paragraph{COVID-19, pneumonia and viral pneumonia:}

\begin{figure}
\centering
\includegraphics[width=0.95\textwidth]{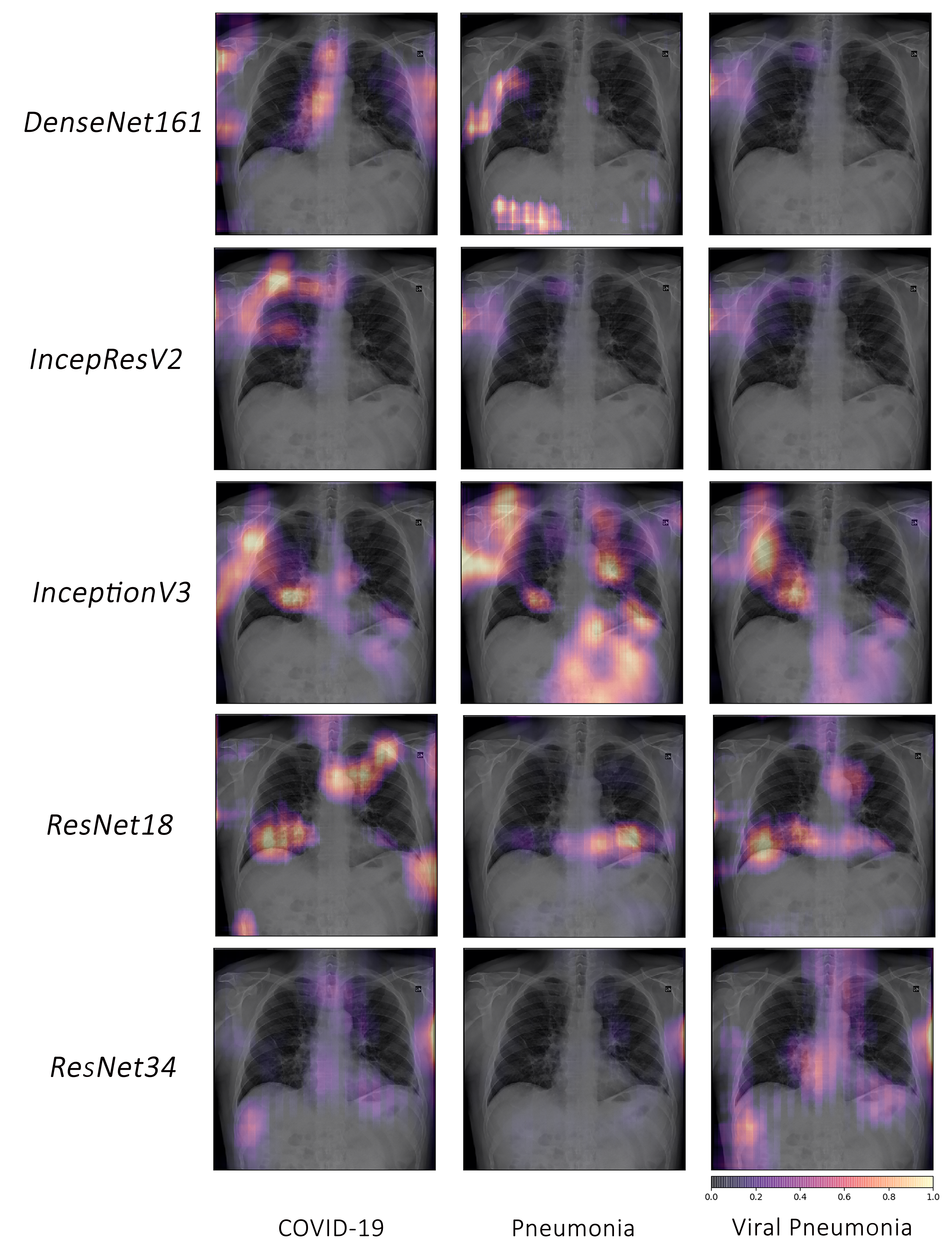}
\caption{Example of occlusion for lung pathologies: COVID-19, pneumonia and viral pneumonia}
\label{fig:Compare_AllModelsOcclusion}
\end{figure}

Based on the fact that COVID-19 is a subset of viral pneumonia, the focus of this section is centralised on the interpretability comparison of the models for these three pathologies. Interpretability techniques reported that different networks focused on different areas for the same CXR image to predict each of the diseases. It was observed that the focus area of DenseNet161 for COVID-19 was explicitly different from that for pneumonia and viral pneumonia. However, InceptionResNetV2 and InceptionV3 emphasised a similar area (different focus areas for each model) for the three pathologies. Furthermore, ResNet18 and ResNet34 targeted the lung region for COVID-19 and viral pneumonia, but differed for pneumonia. Fig~\ref{fig:Compare_AllModelsOcclusion} exhibits the mentioned findings.

\section*{Discussion}

The literature review portrayed that the diagnosis of COVID-19 was seen as a multiclass classification task rather than a multilabel classification. The datasets used in the previous works vary in terms of the amount of data used for the classification task. In \cite{narin2020automatic}, the authors created a balanced dataset by appending the 50 COVID cases with 50 healthy cases from another dataset and reported the highest mean specificity score of 0.90 using InceptionV3. The others~\cite{ zhang2020covid,apostolopoulos2020covid,wang2020covid} performed a multiclass classification task on different imbalanced datasets using X-rays, and achieved a maximum mean specificity of 0.989, 0.979, and 0.971 respectively. In this work, InceptionResNetV2 achieved the highest specificity of 0.975, comparable to previous studies. However, in this research, the authors have used a different dataset, train-test split, and preprocessing techniques compared to previous works, which makes it unfair to compare the results with previous studies.

It was observed that the less complex models were more interpretable, while having fewer dead neurons than the more complex ones. DeneseNet161, which resulted in the highest F1 score, had the highest number of dead neurons and also had the worst focus areas according to interpretability methods. The model that resulted in the second-best F1-score, ResNet18, was the least complex model in this study - while also having the best focus areas as dictated by the interpretability methods. This was further confirmed by a global interpretability method, NAPs, which showed that ResNet18 is less likely to detect biologically irrelevant features. It should be noted that in some cases, the network predicted the findings as a presence of COVID-19, while the doctors did not report any abnormalities.

There were a couple of cases where the network detected both viral and bacterial pneumonia. According to Morris et al.~\cite{denise2017secondaryBactInfect}, and Shigeo et al.~\cite{shigeo2018resViralInduce}, the induction of viral infection could lead to secondary bacterial infection and increase the severity of symptoms. Though such cases were considered as miss-predictions for the current dataset based on the available labels, one could argue that the network was able to detect such instances.


The main motivation to perform a multilabel classification over a multi-class classification was to be able to predict multiple pathologies from the images if they were present. It was observed that all networks, including the Ensemble, were able to correctly predict both COVID-19 and ARDS for the images that had both pathologies present.

Lastly, this study also showed that the models could classify lung pathologies from CXR images, although unwanted objects, such as annotations or labels, were obscuring the radiographs.

\section*{Conclusion and future works}
In this paper, a range of deep learning-based classifiers has been compared for the multilabel classification of COVID-19 and similar pathologies in CXR images, and the interpretability of these models has been investigated and finally corroborated by medical professionals. In general, most of the models performed well. However, certain models failed at specific tasks. The authors have additionally formulated an ensemble employing majority voting, which aided in addressing these models' shortcomings by combining their predictions. Furthermore, the smallest model, ResNet18, was found to compete well with considerably larger models. In fact, for certain situations, it performed better than the largest model in the mix, InceptionResNetV2.  For patients who had more than one pathology, this multilabel classification setup was able to correctly predict all of those pathologies. DenseNet161 was the model that performed the best in this setup in terms of classification scores, though it was observed that the focus of the network was often on unrelated biologically irrelevant regions. This can be attributed to the fact that the network discerned some irrelevant patterns in the dataset, which might be due to the high complexity of the model. The highest number of dead neurons was also observed in this model, suggesting that the model may have been overly complex for the given task. After qualitative analysis of the interpretability results, it can be said that the ResNets were the most interpretable models as the networks predominantly focused on the appropriate regions. 


Model explainability methods such as LIME~\cite{ribeiro2016should}, SHAP~\cite{lundberg2017unified} etc., have not been explored during this research but are planned as future work. The same approach can also be tried on CT images to compare the networks' sensitivity for COVID-19 on CT and CXR images. Moreover, it would be interesting to investigate how the networks' performances are affected if completely unrelated pathologies (like tumours) are mixed with this current dataset. Prior nonimage information (like the patient's prior medical history, the result of the RT-PCR test, etc.) might also be integrated into the network models to aid the networks in decision-making. Furthermore, instead of supplying the whole image to the models, lung segmentation could serve as a preprocessing step, which might improve the networks' predictions by helping them to focus just on the region of interest, which in this case are the lungs. Training techniques such as few-shot learning (including one-shot learning), semi-supervised learning, etc., can be explored for learning to classify COVID-19 cases from a small dataset. Moreover, joint segmentation-classification techniques can also be investigated for this multilabel classification problem. Several interpretation techniques are implemented in the interpretability pipeline, but were not investigated in this study and will be explored in the future for this dataset-model setup. Finally, in the future, a large-scale study involving more medical professionals should also be evaluated to evaluate the benefits of interpretability methods in terms of building trust, and also their usefulness in the clinical workflow should also be evaluated in the future.

\section*{Acknowledgements}
This work was conducted within the context of the International Graduate School MEMoRIAL at Otto von Guericke University (OVGU) Magdeburg, Germany, kindly supported by the European Structural and Investment Funds (ESF) under the programme "Sachsen-Anhalt WISSENSCHAFT Internationalisierung" (project no. ZS / 08/80646) and was partially funded by the Federal Ministry of Education and Research within the Forschungscampus STIMULATE under grant numbers 13GW0473A and 13GW0473B.

\section*{Author contributions statement}
S.C., F.S., C.S., and S.G. created the concept and designed the study, under the supervision of G.R., S.S, O.S., and A.N. S.C., and S.G. performed the experiments. V.K. created the neuron activation profiles and analysed their results. C.S. performed the qualitative analysis of the interpretability results and created the visualisations. R.M. and N.D. reviewed the interpretability results and created the annotations. S.C., F.S., C.S., S.G., and R.K. wrote the manuscript. P.R., G.R., S.S, O.S., and A.N. reviewed and revised the manuscript.

\section*{Data availability}
The datasets generated and/or analysed during the current study are available in the \textbf{COVID-19 image data collection} repository by \textit{Joseph Paul Cohen}, \url{https://github.com/ieee8023/covid-chestxray-dataset} and in the \textbf{Chest X-Ray Images (Pneumonia)} repository by \textit{Paul Mooney}, \url{https://www.kaggle.com/paultimothymooney/chest-xray-pneumonia}.

\section*{Additional information}
All methods were carried out in accordance with the relevant guidelines and regulations. The code of this project is publicly available on GitHub: \url{https://github.com/soumickmj/diagnoPP}.

\bibliography{bibliography}

\end{document}